# On the Detectability of Conflict: a Remote Sensing Study of the Rohingya Conflict

Christopher X. Ren[1], Matthew T. Calef[2], Alice M.S. Durieux[2], A. Ziemann[1], J. Theiler[1]

[1]Intelligence and Space Research Division, Los Alamos National Laboratory, Los Alamos, NM
[2]Descartes Labs, Inc, 100 N Guadalupe St, Santa Fe, NM

**ABSTRACT**

*The detection and quantification of conflict through remote sensing modalities represents a challenging but crucial aspect of human rights monitoring. In this work we demonstrate how utilizing multi-modal data sources can help build a comprehensive picture of conflict and human displacement, using the Rohingya conflict in the state of Rakhine, Myanmar as a case study.*

*We show that time series analysis of fire detections from the Moderate Resolution Imaging Spectroradiometer (MODIS) and Visible Infrared Imaging Radiometer Suite (VIIRS) can reveal anomalous spatial and temporal distributions of fires related to conflict. This work also shows that Synthetic Aperture Radar (SAR) backscatter and coherence data can detect the razing and burning of buildings and villages, even in cloudy conditions. These techniques may be further developed in the future to enable the monitoring and detection of signals originating from these types of conflict.*

## 1. INTRODUCTION

The Rohingya conflict is an ongoing conflict in the northern region of Myanmar's Rakhine State. This conflict is characterized by targeted, sectarian violence against Rohingya Muslim communities and militant attacks by Rohingya insurgents [1]. On the 25th of August 2017, Rohingya insurgents launched coordinated attacks on several police posts and an army base in Rakhine State, resulting in upwards of 70 casualties on both sides [2]. According to a report produced by the Office of the U.N. High Commissioner for Human Rights (OHCHR), released in October 2017, the destruction of Rohingya villages in northern Rakhine State following the events of August 2017 were committed in a '*well-organised, coordinated and systematic*' manner [1]. Information in the report indicates that the dwellings, crops, and livestock of Rohingya Muslims were scorched and destroyed. Following these systematic attacks, an estimated 688000 refugees are estimated to have fled to the Teknaf region of Bangladesh [3], resulting in one of the worst humanitarian crises of the 21st century.

In this work we evaluate the effectiveness of several remote sensing modalities in detecting signals resulting from this conflict. We first utilized fire detections from the Fire information for Resource Management System (FIRMS, derived from satellite observation from NASA's Moderate Resolution Imaging Spectroradiometer (MODIS) [4] and NASA's Visible Infrared Imaging Radiometer Suite (VIIRS) [5]. We also show how synthetic aperture radar (SAR) can be used to detect the razing of villages and construction of facilities following the destruction of settlements.

## 2. FIRE DETECTIONS

MODIS sensors on NASA's Terra and Aqua satellites provide global fire data at a daily temporal resolution and a spatial resolution of ~ 1 km [4], with historical data spanning as far back as 2000. The VIIRS sensor was launched in October 2011 on the Suomi-National Polar-orbiting Partnership (S-NPP) satellite, and provides full global coverage at both 375m and 750 m every 12 hours or less [5]. Both satellites provide active fire products [6], which were used in this study, details of the fire detection algorithms can be found in [4], [5].

In order to evaluate the ability of the FIRMS fire detection products in detecting the fires resulting from the conflict in late 2017, we analyzed yearly trends in the number and location of detections. One of the challenges in this type of analysis is the seasonality present in the number of fire detections. This is largely due to agricultural practices which involve setting old fields on fire in preparation for the upcoming growing season. Biswas *et al.* estimate that between 2003-2012 at the peak of the fire season ( March), the average burnt area was 12900 km$^2$ in Myanmar, with 95000 fire counts from MODIS [7]. In order to account for this highly seasonal behavior, we determine approximate yearly boundaries for the agricultural season by examining the onset and end of the number of fire detections for each year. Figure 2.a shows that each year the fire detections drop to and rises from 0, indicating the beginning and end of the seasons as November to May. In order to detect 'anomalous' behavior on a yearly scale, we then sum the total counts outside of this period (June – October inclusive). These results are shown in Figure 2b, which reveal that 2017 experienced 167 fire detections during this period, or 2.7 standard deviations away from the yearly mean of 29.75 counts. We note that the low number of samples likely results in the mean and standard deviation being severely distorted by this extreme value. The only other year since 2012 which saw over 10 fire detections in this period was 2012, which also saw violence in the region and resulted in an estimated 2500 burnt homes [8].

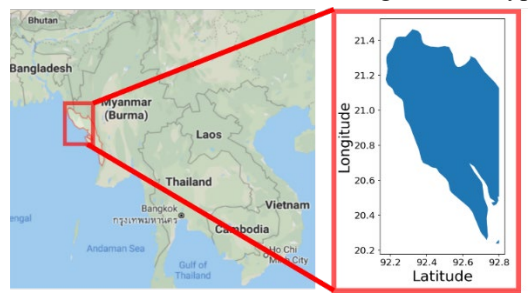

Figure 1: Location of the area of interest, Northern Rakhine in Myanmar.

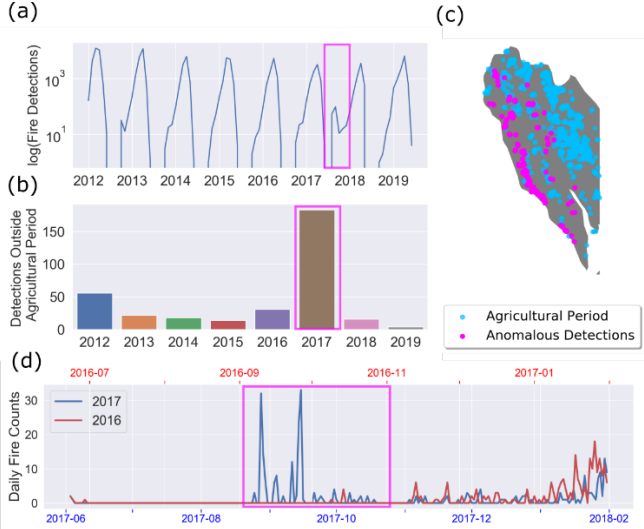

**Figure 2:** a) FIRMS fire detections plotted on a log scale. Note the bump in August/September 2017, highlighted in the magenta box. b) Total fire detections for each year since 2012 in the 'non-agricultural' period (May – December). The counts during this period for 2017 are x3 larger than the closest year. c) Spatial distribution of the August – October 2017 Fires (magenta), compared to those from the Agricultural Season (December – April, in cyan). d) Daily count comparison for the June – February period of the years 2016 and 2017. The counts highlighted in the magenta box correspond to fires occurring simultaneously to reported attacks on Rohingya villages.

Finally, we compare daily fire detections from the June – February periods of 2017, when the violence allegedly occurred, and 2016. The magenta box highlights the source of the 'anomalous' yearly counts for 2017: the period of late August to October shows a sharp, uncharacteristic increase in the number of fires detected over the region, likely corresponding to the systematic and widespread burning of Rohingya homes that was reported during this period [1].

The United Nations Institute for Training and Research (UNITAR) and Digital Globe have provided the locations of totally and partially destroyed settlements, validated by high resolution satellite imagery [9]. Using this data allows us to compare the spatial distribution of our anomalous, out of season fire detections in 2017 to the distribution of the destroyed settlements, as well as the distribution of in-season agricultural fires for that same year. This comparison is shown in Figure 3, where we can see that the distributions of the anomalous fires and destroyed settlements overlap well, relative to the agricultural fires. We can quantitatively establish the difference and similarity between the distributions compute using kernel density estimation (KDE) by computing the mutual information (MI) between the density estimates, given by Eq. 1.

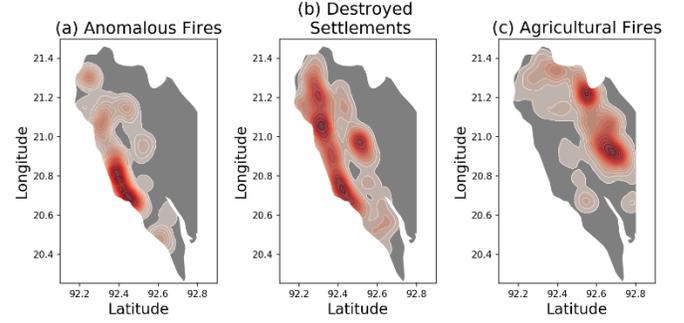

**Figure 3:** KDE plots of the locations of a) the anomalous fire detections, b) the destroyed settlements, c) fires occurring during the agricultural season of 2017.

$$I(X;Y) = \sum_y \sum_x p_{(X,Y)}(x,y) \log\left(\frac{p_{(X,Y)}(x,y)}{p_X(x)p_Y(y)}\right) \quad (1)$$

Where $p_{(X,Y)}(x,y)$ is the joint probability mass function of X and Y, and $p_X(x)$ and $p_Y(y)$ are the marginal probability of X and Y respectively. We use the MI on the density estimates shown in Figure 3 to establish the strength of the relationships between the point patterns exhibited by the anomalous fires, the locations of settlements and the locations of agricultural fires during 2017. These results are shown in Table 1, indicating that the source of the fires is likely to be the destruction of settlements, and that the 'anomalous' fire detections are anomalous both temporally and spatially from fires occurring during the agricultural season.

**Table 1:** MI of the spatial distributions produced by KDE shown in Figure. 3

| I(anomalous; settlements) | I(agricultural; settlements) | I(agricultural; anomalous) |
|---|---|---|
| **0.86** | 0.28 | 0.29 |

## 3. SYNTHETIC APERTURE RADAR

Synthetic Aperture Radar (SAR) can image the surface of the Earth regardless of cloud cover by emitting electromagnetic radiation and quantifying the amplitude of the reflected signal (backscatter), or leveraging the phase information contained in the reflected SAR signal to quantify to phase similarity between pairs of SAR images (interferometric coherence).

We analyzed all Sentinel-1 data acquired between 01/01/2017 and 01/01/2019 over the area of interest in North Rakhine. We generated the interferometric coherence by co-registering Sentinel-1 single-look-complex (SLC) data to a UTM grid, following the "geocoding" approach described in [10]. We chose a grid that aimed to preserve the resolution of Sentinel-1 to the largest degree possible and formed complex interferometric coherence using a Gaussian window.

### 3.1. BACKSCATTER

In this section we evaluate the usage of SAR backscatter in detecting clearance operations conducted in the wake of the August 2017 attacks on Rohingya villages, during which the burnt villages were cleared and demolished. Since the structure of Rohingya villages is typically that of thatched roof houses surrounded by palm trees [11], we use a simple log-ratio based unsupervised change detector [12] to detect clearance operations at the locations of destroyed villages as this methodology has been shown to be effective in detecting deforestation. Figures 4a. and 4b. show the type of change we are attempting to capture in multi-spectral imagery (MSI), though this type of change is not particularly challenging to detect in MSI, this can often be rendered impossible by cloud cover. An effective SAR-based change detection algorithm, which is cloud-penetrating, can thus provide timely detections, and provide a useful monitoring tool.

Our algorithm for change detection over this village is as follows: we first create a median composite of the VH polarized SAR backscatter from 2017/01/01 to 2017/09/01. This aids in reducing the effects of speckle, a common issue in SAR backscatter imagery [13]. With the knowledge that attacks on villages occurred only after late August-September 2017, we use this as our reference image, shown in Figure 4c. We then composite two subsequent backscatter images, and use these as our 'test' image in the log ratio. We then slide this two-image window forwards in time, and take only changes in the log ratio above a threshold of 0.4. This allows us to produce change maps over the area of interest, and detect clearance with high temporal resolution (on average 5 days), as shown in Figure 4e. The total affected area detected is shown in Figure 4d, and correlates well with the affected area shown in the MSI in Figure 4b. We estimate the affected area to be approximately 1300000 $m^2$, and that clearance operations began on 2018/01/20, the date of the first detection.

### 3.2 COHERENCE

Both backscatter and coherence can be useful in distinguishing changes in vegetation: dense vegetation will result in high backscatter but movements in the leaves and stems of vegetation between SAR collections will result in low coherence [14]. Here, we examine changes in coherence observed over the same area shown in Figure 4, where it has been reported that a transit camp has been built to house 8000 refugees [15], visible in Figure 4b. We hypothesize that the transition from thatched roof houses and bare earth following the attacks to concrete and metal structures should lead to a low-to-high coherence change.

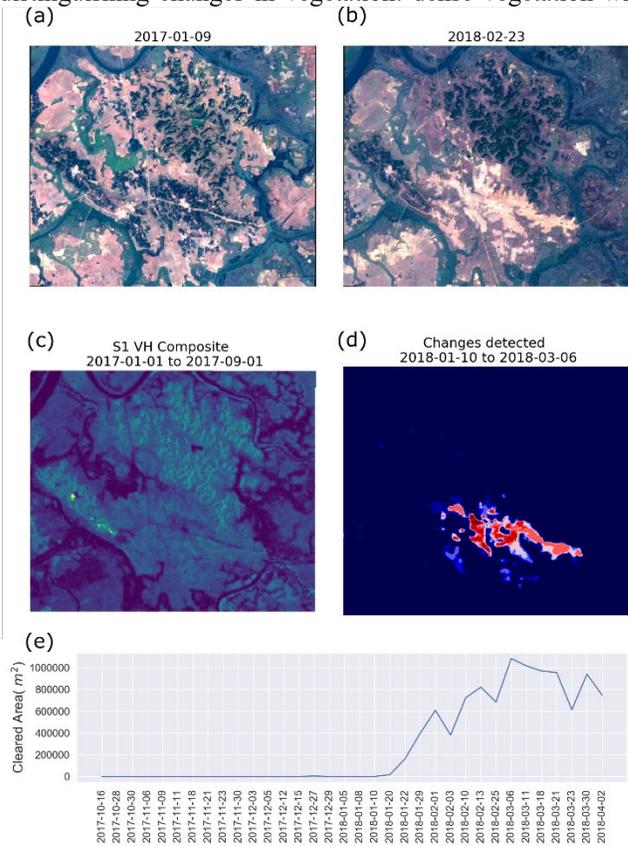

**Figure 4:** a) MSI on 2017/01/01, b) MSI on 2018/02/23 showing that most of the village has been razed/cleared, c) reference VH backscatter composite d) changes detected between 2018/01/20 – 2018/03/06 which correspond to the clearing of the village e) detection area with respect to reference against time.

Figure 5 shows that the coherence between pairs induced by the transit camp built enables us to use simple thresholding to track the construction of the camp regardless of cloud cover. However, this also results in some false positives in the scene as shown in Figure 5d, suggesting a more sophisticated algorithm may be needed. Nonetheless, this demonstrates the utility of InSAR coherence in monitoring these types of events. We also note that producing coherence necessitates a pair of images, thus in change detection we need two pairs of images to capture the change in order to produce a signal in this methodology.

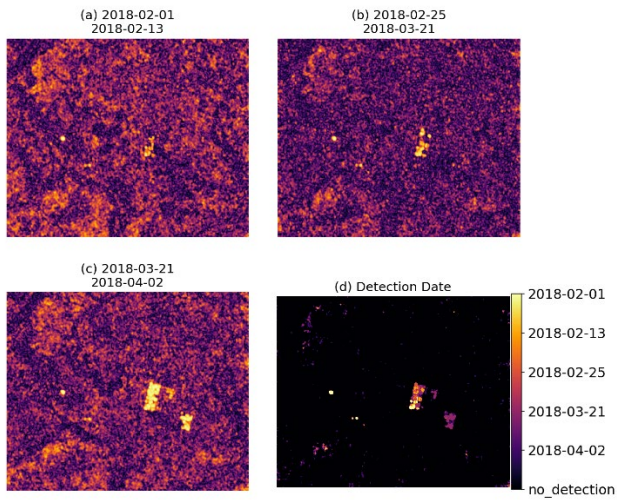

**Figure 5:a), b) and c) Coherence time series tracking the construction of the transit camp. d) detection dates of construction using coherence thresholding.**

### 4. CONCLUSION

We have shown in this paper that many of signals pertaining to the Rohingya conflict in 2017 and their aftermath can be detected using remote sensing modalities. FIRMS fire detections enabled us to establish an anomalous temporal pattern in fire detections. We then compared the spatial distribution of these detections to both ground truth concerning the locations of destroyed settlements and the locations of fires from the agricultural period of 2017, and note a large overlap between the anomalous fires and the locations of the settlements, as quantified by mutual information.

We have also shown that the nature of the Rohingya settlements consisting of thatched roof houses surrounded by trees enables us to track the clearing of entire settlements following the attacks via a SAR backscatter based log-ratio change detection method. Finally we have also shown that SAR interferometric coherence can be used to track the construction of facilities built over the cleared settlements.

The simplistic nature of the algorithms used in this work imply that more sophisticated, targeted methods may be developed using these modalities to provide more accurate, prompt monitoring tools.